# Generating Pairwise Combinatorial Interaction Test Suites Using Single Objective Dragonfly Optimisation Algorithm

Bestoun S. Ahmed

*Software and Informatics Engineering Department, Engineering College, Salahaddin University – Erbil*
*E-mail: bestoun.ahmed@su.edu.krd*

**Abstract**

Combinatorial interaction testing has been addressed as an effective software testing technique recently. It shows its ability to reduce the number of test cases that have to be considered for software-under-test by taking the combinations of parameters as an interaction of input. This combination could be considered as input-configuration of different software families. Pairwise combinatorial test suite takes the interaction of two input parameters into consideration instead of many parameter interactions. Evidence showed that this test suite could detect most of the faults in the software-under-test as compared to higher interactions. This paper presents a new technique to generate pairwise combinatorial test suites. Also, Dragon Fly (DF), a new swarm intelligent optimization algorithm, is assessed. The design and adaptation of the algorithm are addresses in the paper in detail. The algorithm is evaluated extensively through different experiments and benchmarks. The evaluation shows the efficiency of the proposed technique for test suite generation and the usefulness of DF optimization algorithm for future investigations.

**Introduction**

Software quality assurance is one of the important emerging issues in the last decade. It aims to deliver software with minimum defects in which meets specific levels of functionality according to reliability and performance [1]. Besides, it is the systematic pattern of all actions for providing and proving the ability of a software process to build good products [2]. It also tries to improve the development process from the requirement step until the end.

With the vastly growing of software systems and their configurations, there is a chance for fault due to the combinations of these configurations, especially for high configurable software systems. While the traditional test design techniques are useful for fault discovering and prevention, they may not be adequate to take care of faults due to combinations of input components and configurations [3]. Considering all the combinations of configuration leads to exhaustive testing which is impossible due to time and resource constraints [4-6]. For this reason, there is a need to minimize the number of test cases by designing effective test cases that have the same impact as exhaustive testing.

Many sampling strategies have been proposed in the literature. For example, equivalent partitioning partitions the equivalent inputs of the system into groups [7]. The aim here is to design a test suite that covers all of these groups by taking at least one test for each of them. Another example of the sampling strategy is the boundary value where the test suite is designed to cover the boundaries of the input values of the system. Combinatorial interaction testing is another sampling and test design technique that aims to design a test suite to cover all of the combinations among the input parameters of a software system.

Evidence in the literature showed that most of the faults (due to the interaction) in software systems could be detected with pairwise interaction, i.e. when the interaction is between two parameters of the input [8, 9]. Hence, the pairwise strategies aim is to construct pairwise test suites by searching and covering all of the interactions among the input parameters at least once. This could be a difficult task in case of highly configurable systems, which leads to a combinatorial explosion and non-deterministic polynomial-time hard (NP-hard) problems [10]. To this end, different meta-heuristic algorithms have been used to construct pairwise test suites, including Particle Swarm Optimisation (PSO) [11], Genetic Algorithms (GA), [12] Ant Colony Algorithm (ACA) [12], and Tabu Search [13].

Dragonfly algorithm (DFA) [14] is a novel swarm intelligence optimization technique that has been proposed recently. The technique produced promising results as compared to another state of the art techniques and introduced a new algorithm to solve evolutionary style problems. In line with the Search-based Software Engineering (SBSE) practices [15], this paper assesses the use of DFA for pairwise test suite generation. The contribution of the work is mainly twofold: first the paper tries to asses and figures out how DFA benefits the test generation strategy; second, it also shows the design, adaptation, and implementation of this algorithm for test design techniques.

The rest of this paper is organized as follows. Section 2 presents the mathematical notations and objects of combinatorial interaction testing. Section 3 shows the concepts of DFA and discusses its features. Section 4 explains the design concepts of the technique, an adaptation of the algorithm and the implementation. Section 5 contains the results of the evaluation process. Finally, Section 6 concludes the paper.

## Covering Array (Ca)

Covering array (CA) is a combinatorial mathematical object that represents the optimized set of combinatorial test suites. The object assumes that all interactions among the features are occurring within one array. Mathematically, CA is defined as below [16]:

**Definition 1 :** *In its general form, $CA_\lambda(N ; t, k, v)$ is an $N \times k$ array over $(0, \ldots , v - 1)$ such that every $B=\{b_0, ..., b_{d-1}\} \in$ is $\lambda$-covered and every $N \times d$ sub-array contains all ordered subsets from $v$ features of size $t$ at least $\lambda$ times, where the set of column $B=\{b_0, ..., b_{d-1}\} \supseteq \{0, ..., k-1\}$.*

As far as we are looking for the optimal array, the value of $\lambda=1$ which means that all the *t*-tuples of the input functions occur at least once and thus the notation becomes $CA(N;t,k,v)$ [17]. However, this notation assumes that the number of features in each function is equal, which is not the case in real applications. Here, the interaction strength $t=2$ since the pairwise test suite is considered in this paper. Usually, each function has a different number of features depending on the system-under-test. To this end, mixed covering array (MCA) is used as a practical alternative, which is defined as below [18]:

**Definition 2:** *MCA $(N; t, k, (v_1, v_2, \ldots v_k))$, is an $N \times k$ array on $v$ levels, where the rows of each $N \times t$ sub-array cover and all $t$ interactions of values from the $t$ columns occur at least once.*

For more flexibility in the notation, the array can be presented by MCA $(N;d, v^k)$ and can be used for a fixed-level CA, such as CA $(N;t, v^k)$ [18]. So far, these notations have been used within combinatorial testing literature.

## Dragonfly Optimisation

Dragonfly algorithm (DFA) is a novel meta-heuristic algorithm that has been proposed recently by Mirjalili [14]. The algorithm inspired by the natural dynamic and static behavior of dragonflies swarming. The migration behavior inspires the dynamic swarming behavior of the dragonfly and the hunting behavior inspires the static. The former called the "migratory" and the latter called "feeding" swarm. The swarming behavior plays the main role in the inspiration of the algorithm. The static and dynamic behavior of dragonfly is similar to exploration and exploitation in the meta-heuristic optimization algorithms.

To explore the search space, the dragonflies divide themselves to sub-swarms, which is the static swarm that plays the role of the exploration phase. These small groups of dragonflies are flying back and forth in a small area to hunt flying butterflies and mosquitoes. In the dynamic fly, the dragonflies are flying in bigger swarms and in one direction to play the role of exploitation phase [14]. These behaviors of the dragonflies flow three main principles with the neighborhood which are: separation, alignment, and cohesion [19]. Separation refers to the collision avoidance of individuals from others in the neighborhood; alignment arranges the velocity matching among individuals in the neighborhood while the cohesion deals with the tendency of these individuals towards the center of mass in the neighborhood. Figure 1 shows these components of the algorithm with the inspiration of dragonfly behavior.

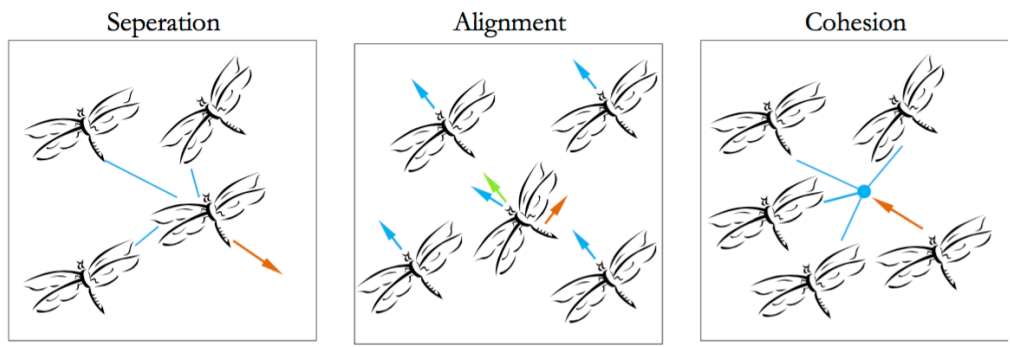

Figure 1. The separation, alignment, and cohesion components of DFA

Since the objective of the swarm is to survive, hence the survival flies toward food source and tries to distract outward enemies [14]. These other two components of DFA are shown in Figure 2.

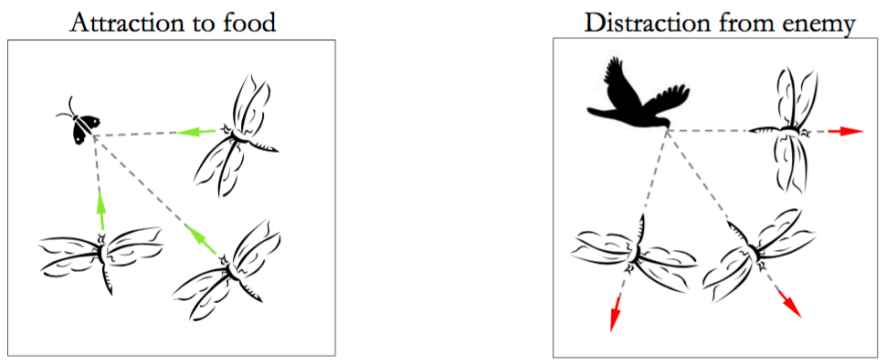

Figure 2. The attraction and distraction of DFA

Mathematically, these five components are represented and modeled as follows. The separation is calculated base on eq.1 [14].

$$S_i = -\sum_{j=1}^{N} X - X_j \ldots \ldots \ldots \ldots \ldots (1)$$

where the number of neighboring individuals denoted as N, the current position of the individual is denoted by X, while $X_j$ shows the position of the $j^{th}$ neighboring individual. The alignment component is calculated base on eq.2.

$$A_i = \frac{\sum_{j=1}^{N} V_j}{N} \ldots \ldots \ldots \ldots \ldots (2)$$

Where the velocity of the $j^{th}$ neighbor is denoted as $V_j$. The cohesion is calculated using Eq. 3

$$C_i = \frac{\sum_{j=1}^{N} X_j}{N} - X \ldots \ldots \ldots \ldots \ldots (3)$$

where the current position of the individual is denoted by X, the number of neighborhoods are denoted as N, and Xj denotes the position of the $j^{th}$ neighbor. The attraction equation can be seen in Eq. 4:

$$F_i = X^+ - X \ldots \ldots \ldots \ldots (4)$$

where $X^+$ is the food source position and X is the current position of the individual. Finally, the distraction of the enemies can be modeled as in eq.5.

$$E_i = X^- - X \ldots \ldots \ldots \ldots (5)$$

where $X^-$ is the enemy position and X is the current position of the individual.

The update role of the algorithm works systematically and artificially by considering two update vectors, Step vector (ΔX) and Position vector (X). DX has an important role in the movement direction of the dragonflies and it is defined in eq.6:

$$\Delta X_{t+1} = (sS_i + aA_i + cC_i + fF_i + eE_i) + w\Delta X_i \ldots \ldots \ldots (6)$$

where t is the iteration counter, (s) is the separation weight, ($S_i$) is the separation of $i^{th}$ individual, (a) is the alignment weight and (A) is the alignment of individual, (c) is the cohesion weight and ($C_i$) is the cohesion of the $i^{th}$ individual, (f) is the food factor and ($F_i$) is the food source of the $i^{th}$ individual, (e) is the enemy factor and ($E_i$) is the enemy position of the individual, and finally (w) is the inertia weight. These factors need accurate running for implementation. In this research we follow the literature [14] to set these parameters, (i.e., w = 0.9–0.2, s = 0.1, a = 0.1, c = 0.7, f = 1, e = 1)

As far as eq.6 calculates this step vector, the vectors' positions are calculated by eq.7 as follows:

$$X_{t+1} = X_t + \Delta X_{t+1} \quad \ldots\ldots\ldots\ldots\ldots\ldots (7)$$

DFA takes the neighborhood as 2D (circle), 3D (sphere), or nD (hypersphere) dimensions of the neighbor dragonflies. This condition is applicable when there is at least one neighbor. However, when the dragonfly cannot find this neighbor, it tends to take a random walk using the Le´vy flight mechanism. Here the position is updated using the Levy update equations [20, 21] as in eq.8:

$$X_{t+1} = X_t + Le'vy(d) \times X_t \quad \ldots\ldots\ldots\ldots (8)$$

where (t) is the current iteration, and d is the dimension of the position vectors. Le´vy flight itself can be obtained in the above equation by eq.9 as follows:

$$Le'vy(x) = 0.01 \times \frac{r_1 \times \delta}{|r_2|^{1/\beta}} \quad \ldots\ldots\ldots\ldots (9)$$

where $r_1$, $r_2$ are random numbers between [0,1], β is a contact variable equals to 1.5. The δ factor in eq.9 is calculated by Eq. 10 as follows:

$$\delta = \left( \frac{\Gamma(1+\beta) \times \sin\left(\frac{\Pi\beta}{2}\right)}{\Gamma\left(\frac{1+\beta}{2}\right) \times \beta \times 2^{\left(\frac{\beta-1}{2}\right)}} \right)^{1/\beta} \quad \ldots\ldots\ldots\ldots (10)$$

where $\Gamma(x) = (x-1)!$

Based on the equations above and arrangements of the DFA, the Pseudo-code of the algorithm is summarised in Algorithm 1.

---

**Algorithm 1:** DFA general algorithm

1. Initialize the dragonflies population $X_i (i = 1, 2, ..., n)$
2. Initialize step vectors $\Delta X_i (i = 1, 2, ..., n)$
3. **while** *the end condition is not satisfied* **do**
4.     Calculate the objective values of all dragonflies
5.     Update the food source and enemy
6.     Update $w, s, a, c, f$, and $e$
7.     Calculate $S, A, C, F$, and $E$ using eq.1 to eq.5
8.     Update neighbouring radius
9.     **if** *a dragonfly has at least one neighbouring dragonfly* **then**
10.         Update velocity vector using eq.6
11.         Update position vector using eq.7
12.     **else**
13.         Update position vector using eq.8
14.     Check and correct the new positions based on the boundaries of variables

## Algorithm Adaptation and Implementation

The DFA is adapted in this research for pairwise test cases generation to assess its optimization efficiency for test generation. Here, the aim is to generate a CA where the interaction strength t=2. The objective function is to cover all pairwise interaction of input parameters at least once. To this end, another algorithm is used to generate the pairwise interactions first. The algorithm iteratively generates these interactions and then put them in a list. This list is used later by the DFA to compute the objective function. Algorithm 2 shows the Pseudo-code of the DFA for pairwise test suite generation.

The DFA starts by taking the input parameters and values. Also, it takes the pairwise interactions list (2-tuples) that must be covered totally. The output of the algorithm is a pairwise test suite that covers the 2-tuples. The algorithm initializes a random dragonfly swarm base on the $k$ columns and filled with $v$ values. Hence, the algorithm iterates until the 2-tuples list get empty. The algorithm evaluates each dragonfly base on its coverage of interactions; then it chooses the best one (dBest). The algorithm then updates the swarm base on the equations 1 to 5. Now, the algorithm evaluates the neighbor dragonflies and updates the velocity and positions base on equations 6 and 7. However, if there is no neighbor dragonfly, the algorithm uses the levy equations for an update as in equation 8. Finally, the best dragonfly is chosen and its corresponding interactions are removed from the 2-tuples.

---

**Algorithm 2:** Dragonfly pairwise generation algorithm

**Input:** Input parameters $k$ and values $v$, all pairwise combinations list $2 - tuples$
**Output:** A pairwise test suite

1. Generate an initial dragonfly swarm base in $k$ and $v$
2. $Iter \leftarrow 1$
3. **while** $2 - tples \neq empty$ **do**
4.     **while** $Iter < Max. Iter$ **do**
5.         **foreach** *dragonfly d* **do**
6.             check coverage with $2 - tuples$ find best best dragonfly "dBest"
7.         Update the food source and enemy
8.         Update $w, s, a, c, f$, and $e$
9.         Calculate $S, A, C, F$, and $E$ using $eq.1$ to $eq.5$
10.        Update neighbouring radius
11.        **if** *a dragonfly has at least one neighbouring dragonfly* **then**
12.            Update velocity vector using $eq.6$
13.            Update position vector using $eq.7$
14.        **else**
15.            Update position vector using $eq.8$
16.     **if** *best coverage is met by* $X_{t+1}$ **then**
17.         $dBest \leftarrow dBest_{t+1}$
18.     Add $dBest$ to the test suite
19.     Remove all the related combinations from the $t - tuples$ list

---

## Evaluation Experiments

The evaluation and benchmarking focuses on the efficiency of the generation for DFA as well as the comparison of the algorithm with otters tools and counterpart algorithms. As far as the efficiency is considered, the size of the final generated test suites (i.e., number of rows) is reported for each algorithm and tool. Here, the benchmarks are adopted from evidence in the literature [11, 22, 23]. The results are compared with the published results for Genetic Algorithm (GA), Ant Colony Algorithm (ACA), and PSO which are heuristic and meta-heuristic counterpart algorithms. Also, the results are also compared with, Jenny [24] IPOG [25], AETG [22] and mAETG [26] which are computational tools. The algorithms are implemented and executed on our environment which consists of Laptop PC with MacOS X El Capitan 10.11.5, 2.9 GHz Intel Core i5, 8GB of DDR3. Tables 1, 2 reported these results. It should be mentioned that cells marked NA (not available) indicate that the results are unavailable from the literature.

Table-1: Comparison of DFA with Existing Published Results

| Configurations | AETG | mAETG | GA | ACA | IPOG | Jenny | TConfig | PICT | PSO | DFA |
|---|---|---|---|---|---|---|---|---|---|---|
| CA (N;2, $3^3$) | NA | NA | NA | NA | 11 | 9 | 10 | 10 | 9 | 9 |
| CA (N;2, $3^5$) | 9 | 9 | 9 | 9 | 12 | 13 | 10 | 13 | 9 | 9 |
| CA (N;2, $3^{10}$) | 15 | 17 | 17 | 17 | 12 | 20 | 20 | 20 | 17 | 17 |
| CA (N;2, $5^{10}$) | NA | NA | NA | NA | 50 | 45 | 48 | 47 | 45 | 45 |
| CA (N;2, $10^{10}$) | NA | NA | 157 | 159 | 176 | 157 | 170 | 170 | 170 | 168 |
| MCA (N;2, $5^1 3^8 2^2$) | 19 | 20 | 15 | 16 | 19 | 41 | 22 | 21 | 21 | 21 |
| MCA (N;2, $6^1 5^1 4^6 3^8 2^3$) | 34 | 35 | 33 | 32 | 36 | 31 | 33 | 38 | 39 | 37 |
| MCA (N;2, $7^1 6^1 5^1 4^6 3^8 2^3$) | 45 | 44 | 42 | 42 | 44 | 51 | 49 | 46 | 49 | 48 |
| MCA(N;2, $10^1 9^1 8^1 7^1 6^1 5^1 4^1 3^1 2^1$) | NA | NA | NA | NA | 91 | 98 | 92 | 101 | 97 | 98 |

Table-2: 10 Parameters Variable Values Comparison of DFA with Existing Tools

| V | IPOG | Jenny | TConfig | PICT | PSO | DFA |
|---|---|---|---|---|---|---|
| 3 | 20 | 19 | 17 | 18 | 17 | 17 |
| 4 | 31 | 30 | 31 | 31 | 29 | 30 |
| 5 | 50 | 45 | 48 | 47 | 45 | 45 |
| 6 | 68 | 62 | 64 | 66 | 62 | 61 |
| 7 | 90 | 83 | 85 | 88 | 81 | 83 |
| 8 | 117 | 104 | 114 | 112 | 109 | 107 |
| 9 | 142 | 129 | 139 | 139 | 139 | 141 |
| 10 | 176 | 157 | 170 | 170 | 170 | 170 |

Tables 1 and 2 show the best-reported results for ten runs of the DFA as well as the comparison between the algorithm and other tools and algorithms. Clearly, from the tables, we can see that DFA generates test suites with satisfactory results as compared to other results. However, as compared to PSO, it generates better results in some cases and equivalent results in others. Overall, in Table 1, GA and ACA generate better results. However, this result is due to the compaction method used to compress the results produced by those algorithms. Hence, we cannot consider those published results as the actual results of the GA and ACA. IPOG, Jenny, PICT, and TConfig can produce impressive results (See Table 2). However, these strategies are designed for specific purposes. For example, IPOG is designed to generate test suites with many parameters and variables and it generates "good enough" in most of the cases. Jenny, PICT, and TConfig are designed to generate test suites faster with. Taking DFA, it can generate test suites efficiently. However, it needs more investigation in case of higher interaction strengths.

## Conclusions

This paper discusses the adaptation, implementation, and evaluation of the new dragonfly algorithm (DFA) for the use of pairwise test suite generation. The algorithm is implemented successfully for this purpose. Experimental evaluation is conducted for the algorithm to evaluate its efficiency against its counterpart PSO algorithm as well as other pairwise tools. The algorithm shows its efficiency for a generation against other algorithms. However, it is not as efficient as expected as compared to PSO. The algorithm shows better results in some benchmarks. However, more evaluation and experiments needed to observe this efficiency. This can be observed clearly by implementing the algorithm for higher combination strengths in the future.